\documentclass[twocolumn,english,aps,arxiv,superscriptaddress]{revtex4-1}
\usepackage{natbib}
\usepackage{grffile}
\usepackage{gensymb}
\usepackage[latin9]{inputenc}
\usepackage{color}
\usepackage{amsmath}
\usepackage{amsfonts}
\usepackage{babel}
\usepackage{graphicx}
\usepackage{epstopdf}
\usepackage{cancel}
\PassOptionsToPackage{normalem}{ulem}
\usepackage{ulem}
\usepackage[unicode=true, bookmarks=false, breaklinks=false,pdfborder={0 0 1},colorlinks=false] {hyperref}
\usepackage{breakurl}

\begin{document}

\title{Magnetic dilution effect and topological phase transitions in (Mn$_{1-x}$Pb$_x$)Bi$_2$Te$_4$}

\author{Tiema Qian}
\affiliation{Department of Physics and Astronomy and California NanoSystems Institute, University of California, Los Angeles,
CA 90095, USA}

\author{Yueh-Ting Yao}
\affiliation{Department of Physics, National Cheng Kung University, Tainan 701, Taiwan}

\author{Chaowei Hu}
\affiliation{Department of Physics and Astronomy and California NanoSystems Institute, University of California, Los Angeles,
CA 90095, USA}

\author{Erxi Feng}
\affiliation{Neutron Scattering Division, Oak Ridge National Laboratory, Oak Ridge,
Tennessee 37831, USA}

\author{Huibo Cao}
\affiliation{Neutron Scattering Division, Oak Ridge National Laboratory, Oak Ridge,
Tennessee 37831, USA}

\author{Igor I. Mazin}
\affiliation{Department of Physics and Astronomy, George Mason University, Fairfax, VA 22030, USA}
\affiliation{Quantum Science and Engineering Center, George Mason University, Fairfax, VA 22030, USA}

\author{Tay-Rong Chang}
\affiliation{Department of Physics, National Cheng Kung University, Tainan 701, Taiwan}
\affiliation{Center for Quantum Frontiers of Research and Technology (QFort), Tainan 701, Taiwan}
\affiliation{Physics Division, National Center for Theoretical Sciences, National Taiwan University, Taipei, Taiwan}

\author{Ni Ni}
\email{Corresponding author: nini@physics.ucla.edu}
\affiliation{Department of Physics and Astronomy and California NanoSystems Institute, University of California, Los Angeles,
CA 90095, USA}

\begin{abstract}

As the first intrinsic antiferromagnetic (AFM) topological insulator (TI), MnBi$_2$Te$_4$ has provided a material platform to realize various emergent phenomena arising from the interplay of magnetism and band topology. Here by investigating (Mn$_{1-x}$Pb$_x$)Bi$_2$Te$_4$ $(0\leq x \leq 0.82)$ single crystals via the x-ray, electrical transport, magnetometry and neutron measurements, chemical analysis, external pressure, and first-principles calculations, we reveal the magnetic dilution effect on the magnetism and band topology in MnBi$_2$Te$_4$. With increasing $x$, both lattice parameters $a$ and $c$ expand linearly by around 2\%. All samples undergo the paramagnetic to A-type antiferromagnetic transition with the N$\acute{e}$el temperature decreasing lineally from 24 K at $x=0$ to 2 K at $x=0.82$. Our neutron data refinement of the $x=0.37$ sample indicates that the ordered moment is 4.3(1)$\mu_B$/Mn at 4.85 K and the amount of the
Mn$_{\rm{Bi}}$ antisites is negligible within the error bars. 
Isothermal magnetization data reveal a slight decrease of the interlayer plane-plane antiferromagnetic exchange interaction and a monotonic decrease of the magnetic anisotropy, due to diluting magnetic ions and enlarging the unit cell. For $x=0.37$, the application of external pressures enhances the interlayer antiferromagnetic coupling, boosting the N$\acute{e}$el temperature at a rate of 1.4 K/GPa and the saturation field at a rate of 1.8 T/GPa. Furthermore, our first-principles calculations reveal that the band inversion in the two end materials, MnBi$_2$Te$_4$ and PbBi$_2$Te$_4$, occurs at the $\Gamma$ and $Z$ point, respectively, while two gapless points appear at $x = $ 0.44 and $x = $ 0.66, suggesting possible topological phase transitions with doping.

\end{abstract}
\date{\today}
\maketitle
\section{Introduction}
Intrinsic magnetic topological insulators provide a great playground for discovering new topological states of matter such as the quantum anomalous Hall insulators, Chern insulators and axion insulators\cite{tokura2019magnetic}. Recently, MnBi$_2$Te$_4$ with the van der Waals bonding was discovered to be the first example of an intrinsic antiferromagnetic (AFM) TI insulator\cite{lee2013crystal,otrokov2019prediction,zhang2019topological,li2019intrinsic, otrokov2019unique}, which has triggered extensive theoretical and experimental studies to explore the emergent phenomena arising from the interplay of magnetism and non-trivial band topology. Soon quantum anomalous Hall effect, Chern insulator state and layer-Hall effect were realized in the two-dimensional (2D) limit of MnBi$_2$Te$_4$\cite{deng2020quantum,liu2020robust, ge2020high,gao2021layer}, opening up great opportunities in low-energy-consumption devices, quantum metrology and quantum computing. 

MnBi$_2$Te$_4$ has a rhombohedral crystal structure with the stacking of Te-Bi-Te-Mn-Te-Bi-Te. The Mn$^{2+}$ ions adopt a high-spin $S = 5/2$ state and order into the A-type AFM structure below 24 K with spins ferromagnetically aligned in-plane and coupled antiferromagnetically along the $c$-axis. It is of particular interest to tune the magnetism and band topology in MnBi$_2$Te$_4$ so that new magnetic topological states and novel functionalities can be realized. Such tuning has been effected by three means.

One is through the structural engineering. Following this line, MnBi$_{2n}$Te$_{3n+1}$ ($n=2, 3, 4$) consisting of alternating $(n-1)$ [Bi$_2$Te$_3$] quintuple layers and one [MnBi$_2$Te$_4$] septuple layer were synthesized \cite{aliev2019novel,147,wu2019natural,1813,klimovskikh2020tunable,ding2020crystal,shi2019magnetic,chen2019topological,lee2019spin,tian2019magnetic,gordon2019strongly}. With increasing $n$, the interlayer Mn-Mn distance increases and thus the AFM interlayer exchange interaction decreases. Consequently, MnBi$_2$Te$_4$, MnBi$_4$Te$_7$ and MnBi$_6$Te$_{10}$ are Z$_2$ AFM topological insulators while MnBi$_8$Te$_{13}$ becomes a ferromagnetic axion insulator \cite{1813}. 

Another is through external pressure \cite{chen2019suppression,shao2021pressure, qian2022unconventional}, where pressure-activated metamagnetic transitions \cite{qian2022unconventional} were reported. The third approach is chemical doping. When Sb is doped in MnBi$_2$Te$_4$\cite{chen2019intrinsic,yan2019evolution}, Sb not only substitutes Bi, but also leads to complex chemical disorders. Due to the similar ionic radius between Mn$^{2+}$ and Sb$^{3+}$, the amount of Mn on the Mn site (Mn1 sublattice) decreases while the amount of the Mn$_{\rm{Bi, Sb}}$ antisites, that is, the amount of the Mn on the Bi/Sb site (Mn2 sublattice) increases \cite{liu2021site, hu2021tuning}. Consequently, holes are doped into the system,
and the ground state becomes ferrimagnetic with decreasing saturation moment and saturation field \cite{murakami2019realization, lai2021defect}. Therefore, the uncontrollable and complex chemical disorders caused by Sb doping make it challenging to differentiate the effect caused by the dilution of the Mn1 sublattice and the growing of the Mn2 sublattice. To investigate the effect of magnetic dilution of the Mn1 sublattice on the magnetism and band topology, here we grew and characterized (Mn$_{1-x}$Pb$_x$)Bi$_2$Te$_4$ $(0\leq x \leq 0.82)$ single crystals. We find that the  Mn$_{\rm{Bi}}$ antisites remain negligible. We show that the dilution of the Mn1 sublattice leads to linearly  decreasing with doping N$\acute{e}$el temperature and saturation field. We further reveal a complicated band inversion evolution upon doping, where two gapless points appear when doping concentration achieves $x = $0.44 and 0.66.

\section{methods}

Single crystals of (Mn$_{1-x}$Pb$_x$)Bi$_2$Te$_4$ were grown using the self flux method \cite{hu2021growth}. Pb shots, Mn pieces, Bi and Te chunks were mixed with a ratio of
[$x_\mathrm{nominal}$Pb$ + (1-x_\mathrm{nominal}$) Mn]Te : Bi$_2$Te$_3$ varying from 15 : 85, 21 : 79, 29 : 71, 31 : 69, 37 : 63 and 30 : 70 for $x_{\rm{nominal}}=$ 0, 0.36, 0.5, 0.6, 
0.7 and 0.85. The mixture was loaded into an alumina crucible and vacuum sealed inside a quartz tube. It was then heated to 900 $\degree$C in 4 hours and cooled to 598 $\degree$C in 0.5 hours. Then the ampule was cooled from 598$\degree$C to 592 $\degree$C in a duration of 3 days and stayed at 592 $\degree$C for 3 more days. The ampule was then centrifuged and shiny single crystals with lateral sizes $\sim$ 3$\times$3 $mm^{2}$ can be obtained.

\begin{table*}[]

\setlength{\tabcolsep}{2.4pt}
\caption{Data summary of (Mn$_{1-x}$Pb$_x$)Bi$_2$Te$_4$. $x$ refers to the molar ratio of Pb/(Pb+Mn) obtained by the WDS measurements. The lattice parameters obtained by PDXR refinement ($a$ and $c$ in \AA). The effective magnetic momentum ($\mu_{\rm{eff}}$ in $\mu_B$/Mn) and Curie-Weiss temperature ($T_{CW}$ in K) are calculated from Fig. \ref{T-dependence} (see text). Saturation moment ($\mu_{\rm{s}}$ in $\mu_B$/Mn) for $x=0$ (2 K, 7.7 T) and $x >0$ (2 K, 7 T), effective magnetic anisotropy ($SK$ in meV) and effective interlayer magnetic interaction ($SJ_c$ in meV) are obtained from magnetization measurements shown in the first row of Fig. \ref{H-dependence} (see text), charge carrier density ($n$ in $10^{20} cm^{-3} $) is calculated from Hall measurements shown in the third row of Fig. \ref{H-dependence} (see text). }

\label{table:WDS}
\begin{tabular}{c|c|c|c|c|c|c|c|c|c|c|c}
\hline
$x_{\mathrm{nominal}}$ & WDS& $x $&$a$&$c$& $T_N$&$T_{CW}$&$\mu_{\mathrm{eff}}$& $\mu_\mathrm{s}$ & $SK$& $SJ_c$ &$n$\\\hline
0        & Mn$_{0.88(1)}$Bi$_{2.08(1)}$Te$_4$  & 0 & 4.3314(2)&40.915(4) &23.0& 5.0& 5.4  & 4.5&0.080&0.090 &  1.3\\\hline
0.36      & Mn$_{0.64(1)}$Pb$_{0.16(1)}$Bi$_{2.16(2)}$Te$_4$ & 0.20(1)&4.3560(6)&41.05(2) &18.0 &5.0& 5.5  & 4.4&0.035&0.065&2.8\\\hline
0.5      & Mn$_{0.55(4)}$Pb$_{0.33(4)}$Bi$_{2.10(2)}$Te$_4$ & 0.37(3)&4.3763(3)&41.201(7)&14.5 & 6.5& 5.7 & 4.5&0.030&0.055&3.5\\\hline
0.6      & Mn$_{0.38(1)}$Pb$_{0.43(1)}$Bi$_{2.19(1)}$Te$_4$ & 0.53(2)&4.3916(7)&41.33(1)&9.5 & 4.0& 6.0  & 4.7&0.030&0.040&4.1\\\hline
0.7      & Mn$_{0.24(1)}$Pb$_{0.55(4)}$Bi$_{2.15(1)}$Te$_4$ & 0.69(4)&4.407(1)&41.44(1) &4.5& 2.5& 5.7& 4.6&0.025&0.025&12.9\\\hline
0.85     & Mn$_{0.14(1)}$Pb$_{0.67(1)}$Bi$_{2.20(1)}$Te$_4$ & 0.82(4)&4.4210(5)&41.56(1) &2.0& 0& 5.7 & 4.7 & $-$ & $-$ & 24.6 \\\hline
\end{tabular}
\end{table*}

Extra care was paid to select for measurements the purest specimens without any detectable impurities. X-ray diffraction was performed
using a PANalytical Empyrean diffractometer (Cu K$\alpha$). Initially, the surface x-ray diffraction was used in order to select a specimen of the right phase, then a small portion of the latter was grounded for powder x-ray (PDXR) diffraction to check for impurities. Once no discernible impurity peaks were detected, the same specimen was then used for all measurements. First, chemical analysis was performed using the wavelength-dispersive spectroscopy (WDS). Hereafter, the doping concentration $x$ refers to the molar ratio of Pb/(Pb+Mn) determined by the WDS measurements. Then, the magnetization data were measured in a Quantum Design Magnetic Properties Measurement System (QD MPMS3). Lastly, the sample was cleaved into thin plates and cut into rectangular bars for six-probe electric and Hall resistivity measurements, which were performed in a QD DynaCool Physical Properties Measurement System (QD PPMS).

Transport measurements under pressure were performed on the $x = 0.37$ sample. Hydrostatic pressure was applied in a high pressure cell designed by C\&T factory, compatible with the QD PPMS. Daphne Oil 7373 was used as the hydrostatic pressure medium.  

Single-crystal neutron diffraction was performed for the $x = 0.37$ sample at 4.85 K and 0 T on the HB-3A DEMAND single-crystal neutron diffractometer located at Oak Ridge National Laboratory\cite{chakoumakos2011four}. 

The bulk band structures of PbBi$_2$Te$_4$ and MnBi$_2$Te$_4$ were computed using the projector augmented wave method as implemented in the VASP package \cite{blochl1994projector,kresse1999ultrasoft,kresse1996efficiency} within the GGA \cite{perdew1996generalized} and GGA plus Hubbard U (GGA + U) \cite{dudarev1998electron} scheme, respectively. On-site $U = 5.0$ eV was used for Mn $d$ orbitals. The spin-orbit coupling was included self-consistently in the calculations of electronic structures with a Monkhorst-Pack k-point mesh 15 $\times$ 15 $\times$ 5. The experimental lattice parameters were used. The atomic positions were relaxed until the residual forces were less than 0.01 eV/\AA. In order to simulate the doping effect, we constructed a tight-binding Hamiltonian for both PbBi$_2$Te$_4$ and MnBi$_2$Te$_4$, where the tight-binding model matrix elements were calculated by projecting onto the Wannier orbitals \cite{marzari1997maximally,souza2001maximally,mostofi2008wannier90}, which used the VASP2WANNIER90 interface \cite{franchini2012maximally}. We used Bi $p$ orbitals and Te $p$ orbitals to construct Wannier functions, without performing the procedure for maximizing localization. The electronic structures of the doped compounds were then calculated by a linear interpolation of tight-binding model matrix elements of the Hamiltonians.

\begin{figure}
    \centering
    \includegraphics[width=3.3in]{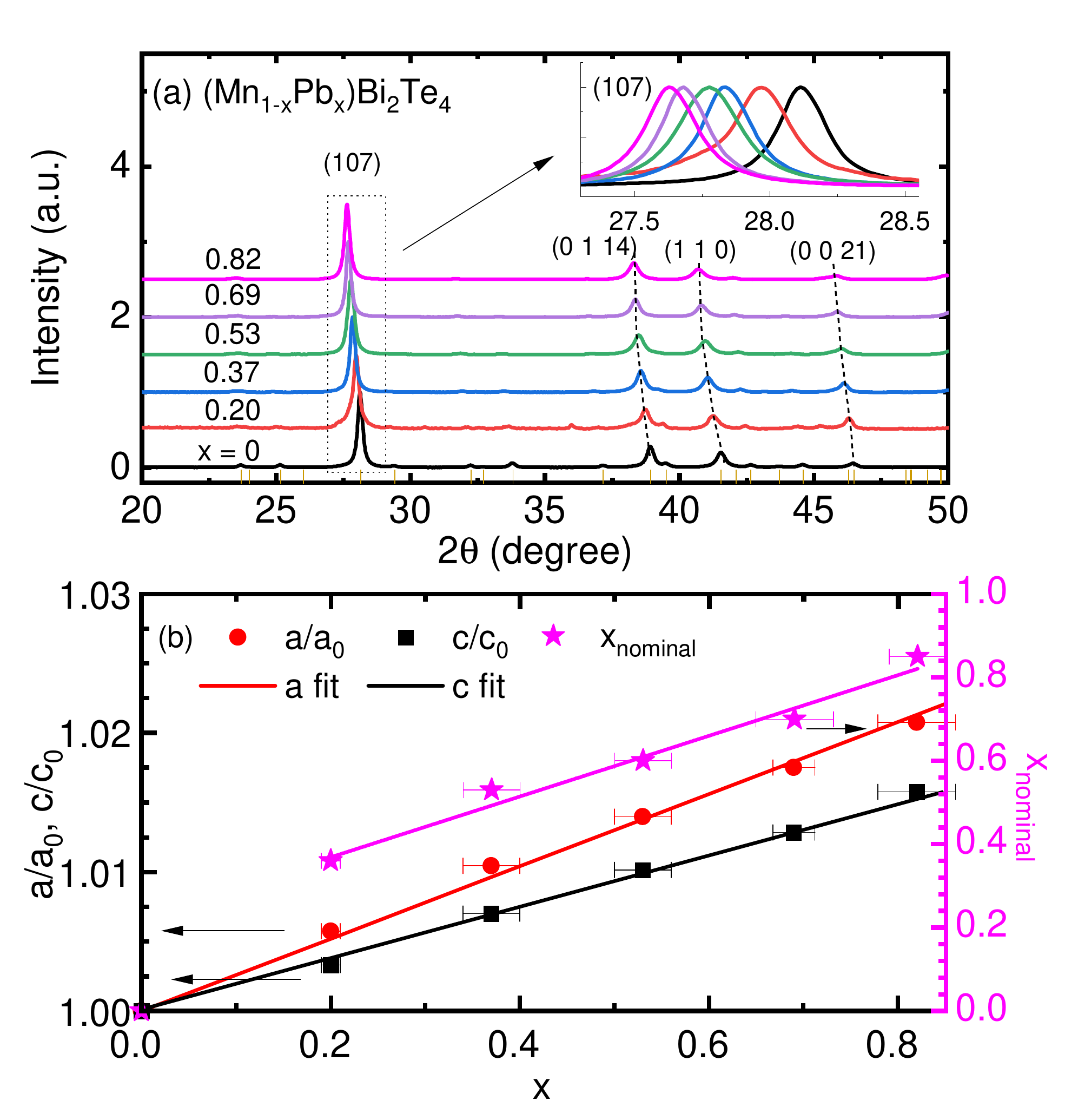}
    \caption{X-ray diffraction of (Mn$_{1-x}$Pb$_x$)Bi$_2$Te$_4$. (a) PXRDs for various concentrations. Inset: the zoom-in plot of the (1 0 7) PXRD peak. (b) The doping-dependent relative lattice parameters $a/a_{0}$, $c/c_{0}$ and nominal concentration used in growth. $a_{0}$ and $c_{0}$ are the lattice parameters of MnBi$_{2}$Te$_{4}$.}
    \label{x-ray}
\end{figure}

\section{Results} 
\begin{figure*}
    \centering
    \includegraphics[width=7in]{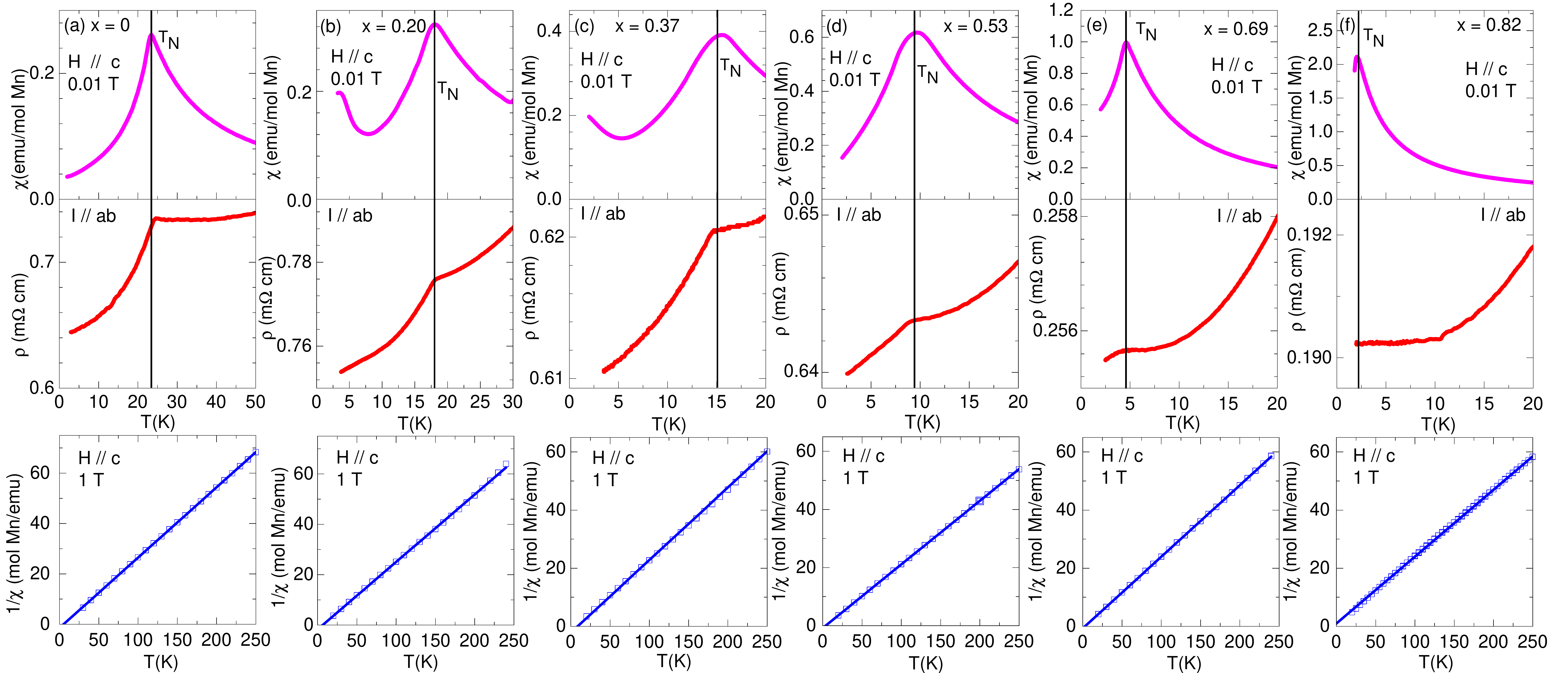}
    \caption{The evolution of temperature-dependent properties of (Mn$_{1-x}$Pb$_x$)Bi$_2$Te$_4$. Top row: $\chi(T)$, the temperature-dependent magnetic susceptibility under 0.01 T with $H \parallel$ c. Middle row: $\rho_{xx}(T)$, the temperature-dependent electrical resistivity with the current along the $ab$ plane. Ordering temperature $T_N$ for each concentration is marked and a correlation between two measurements can be observed. Bottom row: $1/\chi(T)$, the inverse magnetic susceptibility measured at 1 T above $T_N$. Curie-Weiss fits are shown in solid lines. }
    \label{T-dependence}
\end{figure*}

Both WDS and PXRD measurements indicate that Pb successfully substitutes Mn in MnBi$_2$Te$_4$. The results are summarized in Fig. \ref{x-ray} and Table \ref{table:WDS}. Figure \ref{x-ray}(a) shows the PDXR for various doping levels. All the peaks can be indexed by the MnBi$_2$Te$_4$ phase. If there is Bi$_2$Te$_3$ impurity, an additional hump can be seen on the right shoulder of the (107) peak. As shown in the inset of Fig.\ref{x-ray}(a), the Bi$_2$Te$_3$ phase is almost indiscernible. According to Table. \ref{table:WDS} , in MnBi$_2$Te$_4$, the molar concentration of (Mn+Pb) is 0.88(1) while the molar concentration of Bi is 2.08(1). This is consistent with the neutron and x-ray studies which reveal the partial occupancy of Bi atoms on the Mn sites. Upon doping, the amount of (Mn+Pb) stays around 0.80 while the amount of Bi is between 2.1 and 2.2, providing strong evidence that indeed Pb substitutes Mn atoms, not Bi. As plotted in Fig. \ref{x-ray}(b), the real doping level $x$ defined as Pb/(Pb+Mn) from the WDS data increases with the nominal doping level $x_{\rm{nominal}}$. From 0.2 to 0.82, a linear fitting results in $x=-0.28+1.34x_{\rm{nominal}}$. Figure \ref{x-ray}(b) also shows the evolution of lattice parameters with respect to $x$. The lattice parameters $a$ and $c$ increase linearly by 2.2$\%$ and 1.8$\%$ respectively from $x=0$ to 0.82, consistent with Vegard's law. This is different from the Sb-doped MnBi$_2$Te$_4$ where $c$ remains unchanged but $a$ decreases with doping. 

\begin{figure*}
    \centering
    \includegraphics[width=7in]{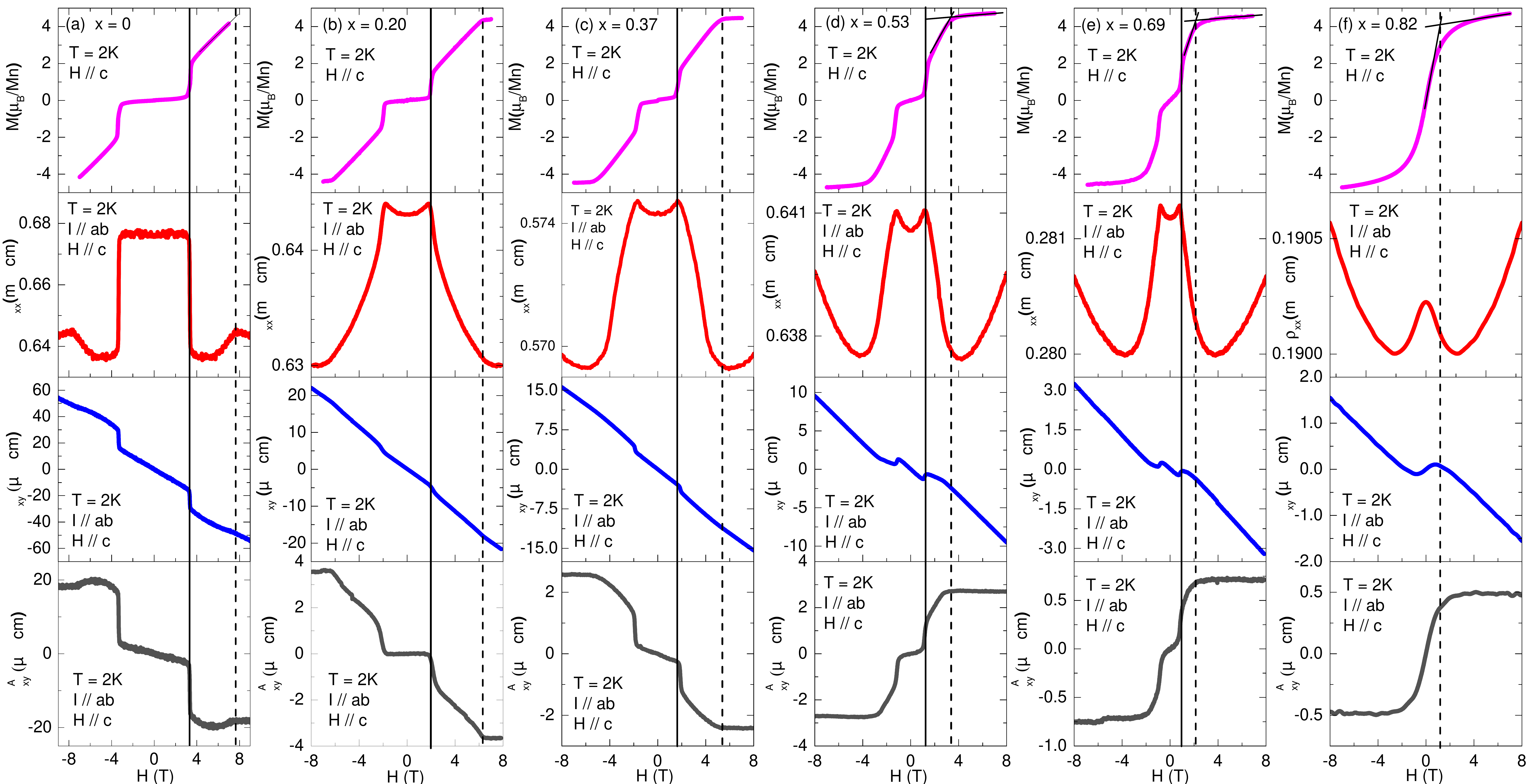}
    \caption{The evolution of magnetic-field-dependent properties of (Mn$_{1-x}$Pb$_x$)Bi$_2$Te$_4$. First row: $M(H)$, the isothermal magnetization at 2 K with $H \parallel c$. The reflection-point criterion used to determine the $H_s$ is shown for $x\geq3$. Second row: $\rho_{xx}(H)$, the magnetic field dependence of electrical resistivity with the current along the $ab$ plane and $H \parallel c$. Third row: $\rho_{xy}(H)$, the Hall resistivity with the current along $ab$ plane and $H \parallel c$. Forth row: $\rho_{xy}^{A}(H)$, the anomalous Hall resistivity calculated by subtracting linear Hall background in $\rho_{xy}(H)$.}
    \label{H-dependence}
\end{figure*}

\subsection{Magnetic and electrical transport properties}
Magnetic and electrical transport properties of this doping series are shown in Figs. \ref{T-dependence} and \ref{H-dependence}. The evolution of the magnetism throughout the doping process can be well traced in the temperature-dependent susceptibility with $H\parallel c$ ($\chi(T)$) and the temperature-dependent resistivity with $I\parallel ab$ ($\rho_{xx}(T)$) in Fig. \ref{T-dependence}. For $x = 0$, a sharp cusp in $\chi(T)$ and a drop in $\rho_{xx}$ agree with the previous reports, indicating a paramagnetic (PM) to A-type AFM phase transition at $T_{N}$ = 24 K. 
The cusp feature in $\chi(T)$ persists for $x\leq 0.82$ while the drop in $\rho_{xx}$ can be observed up to $x=0.69$. Together with the small magnitude of $\chi(T)$ across the whole doping series, these observations indicate the A-type AFM ground state with $T_N$ decreasing monotonically from 24 K for $x=0$ to 2 K for $x=0.82$. We note the drop in $\rho _{xx}$ at $T_N$ becomes less dramatic upon doping (indiscernible at $x=0.82$), which is consistent with the fact that the fewer the magnetic scattering centers, the weaker the spin disorder scattering.

The bottom panel of Fig. \ref{T-dependence} presents the inverse magnetic susceptibility, $1/\chi$, with $H\parallel c$ and $H=1$ T. As one can see, $1/\chi$ is rather linear in essentially entire range between 40 K to 250 K. The Curie-Weiss fitting results in the effective moment $\mu_{\rm{eff}}$ of 5.7$\pm0.3$ $\mu_B$/Mn with no clear doping-dependence (Table I). This is consistent with the theoretical value of 5.9 $\mu_B$/Mn for high-spin Mn$^{2+}$. The Curie-Weiss temperature $T_{CW}$ is positive for $x \leq 0.69$, consistent with the strong in-plane FM fluctuation; $T_{CW}$ becomes 
zero at $x=0.82$, suggesting AFM spin fluctuation in the paramagnetic state, likely due to the Mn lattice being very dilute.

Figure \ref{H-dependence} presents the $M(H)$ (isothermal magnetization), $\rho_{xx}(H)$,  $\rho_{xy}(H)$ (Hall resistivity) and $\rho^A_{xy}(H)$ (anomalous Hall resistivity) with $H \parallel c$ at 2 K. Except for the $x=0.82$ sample, where no spin-flop feature appears at 2 K, all other samples with $x \leq 0.69$, a spin-flop transition can be well resolved in $M(H)$. The spin-flop transition field $H_{\rm{sf}}$ marked by the vertical line decreases with increasing Pb doping, from 3.3 T for $x=0$ to 0.92 T for $x=0.69$. Meanwhile, the saturation field $H_s$ marked by the vertical dash line also decreases with $x$, from 7.7 T \cite{yan2019evolution} for $x=0$ to 2.1 T for $x=0.69$ and 1.2 T for $x=0.82$. Furthermore, unlike Sb-doped MnBi$_2$Te$_4$ where the saturation moment decreases to 2.0 $\mu_B$/Mn for MnSb$_2$Te$_4$ due to the formation of $\sim$ 16\% of Mn$_{\rm{Bi}}$ antisites\cite{yan2019evolution, frachet2020hidden}, in all Pb-doped MnBi$_2$Te$_4$ samples, the magnetic moment at 7 T and 2 K remains around 4.5 $\mu_B$/Mn (Table. I). This provides strong evidence that the amount of such antisites remains minimal during Pb doing.

Despite Bi and Te dominating the band characters at the Fermi level and the Mn band being a few eV away from the Fermi level, charge transport strongly couples to the magnetism. As shown in the second row in Fig. \ref{H-dependence}, at $x = 0$, upon increasing the field, a sharp decrease of $\rho_{xx}$ happens at $H_{\rm{sf}}$ due to the loss of spin-disorder scattering when the system goes from the AFM state to the canted AFM state; $\rho_{xx}$ then slightly increases in the canted AFM state and reaches a kink feature at $H_s$. A negative slope of $\rho_{xy}$ indicates the electrons dominate the charge transport while the $\rho^A_{xy}(H)$ shows a sharp drop at $H_{\rm{sf}}$ and becomes independent to the $M(H)$ in the canted AFM state.
Upon doping, electrons remain the dominant carrier in charge transport, which is in stark contrast with the Sb-doped MnBi$_2$Te$_4$, again suggesting the amount of the Mn$_{\rm{Bi}}$ antisites remains few. The sharp drop from both $\rho_{xx}$ and $\rho_{xy}$ continues to appear at $H_{\rm{sf}}$ for $x \leq 0.69$. As shown by the solid lines, the $H_{\rm{sf}}$ from three measurements corresponds well with each other. We can determine $H_{s}$ using $M(H)$ and $\rho^A_{xy}(H)$, indicated by the dash lines. For $x=0.82$, at 2 K where it just orders, no feature signaling $H_{\rm{sf}}$ can be observed while the $H_{s}$ can be consistently determined by both $M(H)$ and $\rho^A_{xy}$ using the criterion shown in the first row of Fig. \ref{H-dependence}. 

\begin{figure}
    \centering
    \includegraphics[width=3.3in]{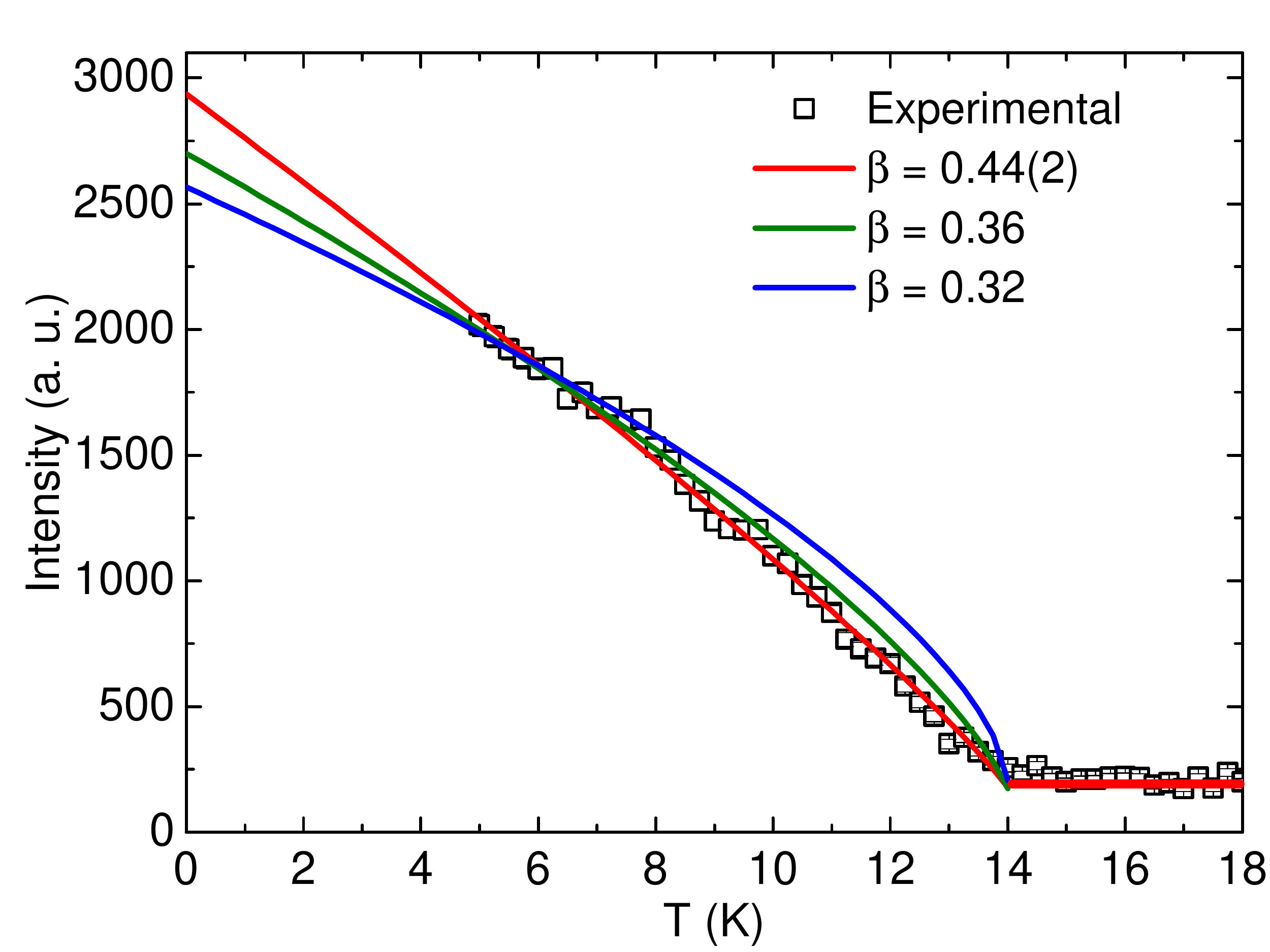}
    \caption{The temperature-dependent neutron peak intensity at magnetic reflection (1 0 -0.5) for the $x = 0.37$ sample. Order parameter fit results in a value of $\beta=0.44(2)$. Curves with $\beta=0.36$ (3D Heisenberg case) and $\beta=0.32$ (3D Ising case) are also shown for comparison.}
    \label{orderparameter}
\end{figure}

To further investigate the crystal and magnetic structures, single-crystal neutron diffraction was performed on the $x=0.37$ sample at 4.85 K. The refined structural parameters are summarized in Table II. Since Mn, Bi and Pb co-occupy the Mn site ($3a$ site), which complicates the refinement, to make the refinement work, we confined the Pb concentration as the one obtained from the WDS measurement. Meanwhile, if we allow the (Mn, Pb)$_{\rm{Bi}}$ antisite formation, that is, if we allow Mn and Pb to partially occupy the Bi site ($6c$ site) in the refinement, the obtained Bi concentration is too low to agree with the WDS measurement, suggesting that the amount of (Mn, Pb)$_{\rm{Bi}}$ antisites is negligible within the neutron measurement resolution. Our refinement leads to the chemical formula of Mn$_{0.50(1)}$Pb$_{0.33(1)}$Bi$_{2.17(1)}$Te$_4$, which agrees well with the WDS values. Using the crystal structural information, the refinement of the magnetic Bragg peaks results in an ordered moment of 4.3(1) $\mu_B$/Mn at 4.85 K.

Figure \ref{orderparameter} shows the peak intensity of the magnetic reflection (1 0 -0.5). It follows an empirical power law behavior, 
\begin{equation}
I = A \left(\frac{T_{N} - T}{T_{N}}\right)^{2\beta} + B
\end{equation}
where $A$ is a proportional constant, $\beta$ is the critical exponent of the order parameter, and $B$ is the background. Unlike the undoped sample whose order parameter can be fitted by the 3D Heisenberg model near the critical temperature \cite{ding2021neutron}, from 5 to 20 K, the best fit is shown as the red curve, which yields $T_N$ = 14.1 K and the critical exponent $\beta$ = 0.44(2), considerably larger than that in MnBi$_2$Te$_4$ (0.36)\cite{ding2021neutron}. We also show the curves with $\beta$ = 0.36 (3D Heisenberg case) and $\beta$ = 0.32 (3D Ising case), which clearly deviate from the data. Note that $\beta=0.44$ is very close to the mean-field value, 0.5, and cannot represent the true 
criticality in any sensible Hamiltonian (nor do we expect the Hamiltonian class to change with doping). On the other hand, this number 
is rather close to critical exponents expected in various percolation models \cite{christensen2002percolation}. Thus, the temperature evolution of the observable order parameter may reflect static percolation, expected in this strongly disordered medium, rather than dynamic fluctuations.

\begin{figure}
    \centering
    \includegraphics[width=3.4in]{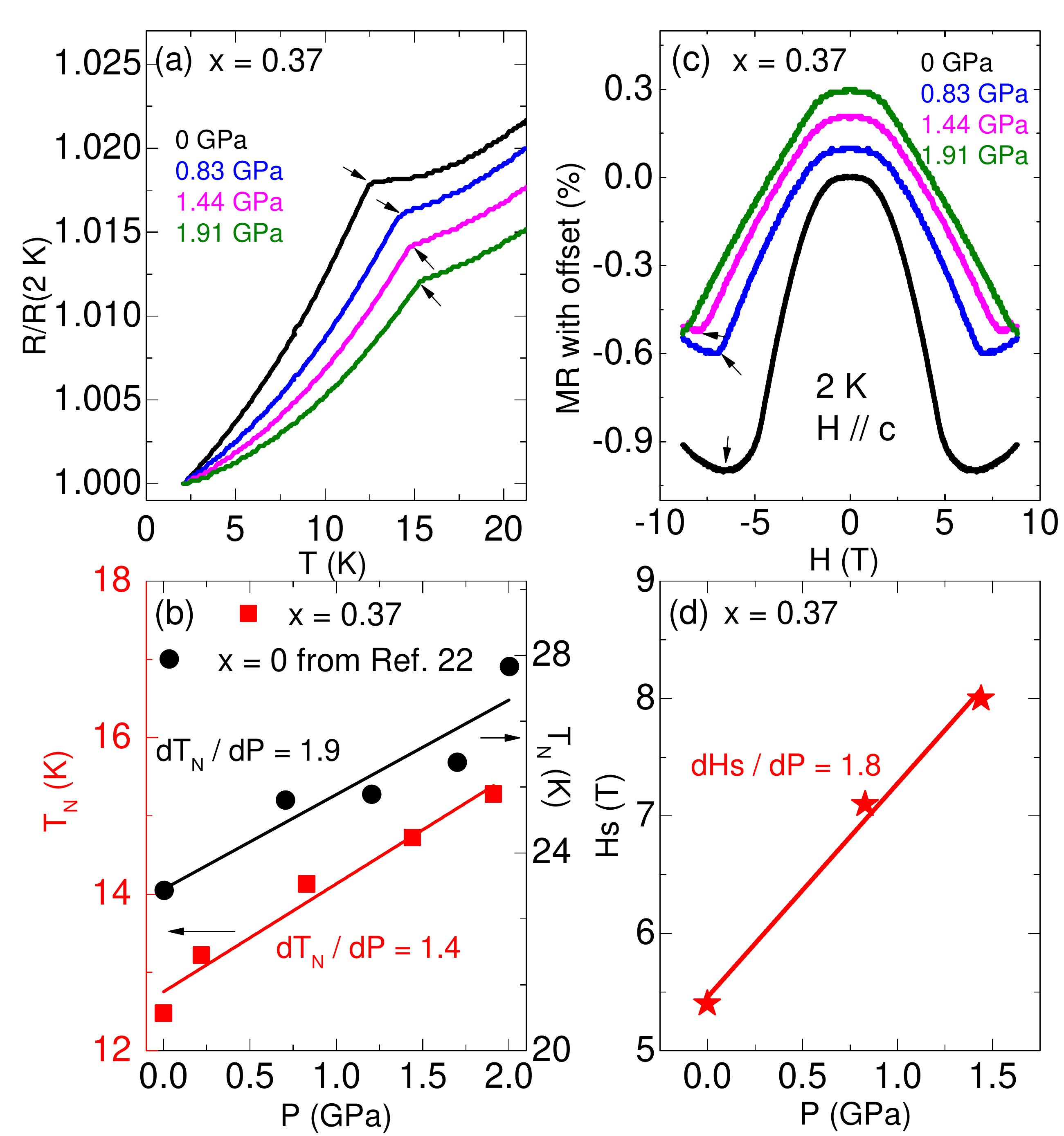}
    \caption{Pressure measurement of the $x=0.37$ sample. (a) The temperature dependence of $\rho_{xx}$ at different pressures. $T_N$ is marked by black arrows. (b) The evolution of $T_N$ with pressure. Linear fits of both data were shown in lines. (c) The field dependence of $\rho_{xx}$ at different pressures with offset. $H_s$ is marked by black arrows. (d) The evolution of $H_{s}$ with pressure. Linear fit is shown in line.}
    \label{Pressure}
\end{figure}

\begin{table}[]
\setlength{\tabcolsep}{3pt}
\label{table:Neutron}
\renewcommand{\arraystretch}{1.5}
\caption{Refined structural parameters for the $x=$0.37 sample based on the single crystal neutron diffraction data. (number of reflections: 192; $R_F = 3.83\%$; $\chi^2 = 28.7$)}
\centering
\begin{tabular}{ccccccc}
\hline
Atom & site & $x$     & $y$     & $z$          & occ.    & Moment at 4.85 K          \\
\hline\hline
Mn1  & $3a$& 0     & 0     & 0         & 0.50(1)         & 4.3(1) $\mu_{B}$/Mn   \\
Bi1  & $3a$&0     & 0     & 0        & 0.17(1)                     \\
Pb1  & $3a$&  0   & 0     & 0        & 0.33(1)               \\
Bi2  & $6c$&0 & 0 & 0.42645(4)   & 1               \\
Te1  &$6c$& 0     & 0     & 0.13459(6)          & 1                  \\
Te2  &$6c$& 0 & 0 & 0.29202(5)   & 1                 \\
\hline\hline
\end{tabular}
\end{table}

\begin{figure*}
    \centering
    \includegraphics[width=7in]{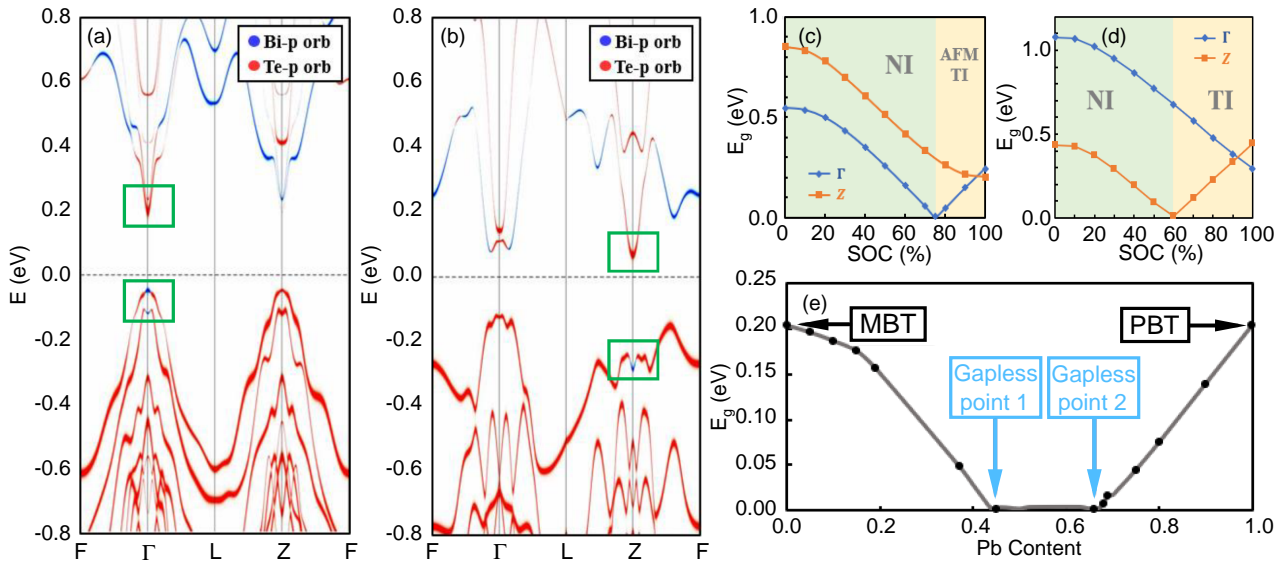}
    \caption{(a) The bulk band structure of MnBi$_2$Te$_4$ and (b) PbBi$_2$Te$_4$. The blue and red dots indicate the weight of the Bi-$p$ orbitals and Te-$p$ orbitals, respectively. (c) The gap value at $\Gamma$ and $Z$ point as a function of SOC strength for MnBi$_2$Te$_4$ and (d) PbBi$_2$Te$_4$. (e) The minimum bulk band gap along $\Gamma$ - $Z$ as function of Pb content.}
    \label{DFT}
\end{figure*}

To further study how interlayer and intralayer interactions will affect the magnetism in the doped samples, we measured the $x=0.37$ sample under different hydrostatic pressure. Figure \ref{Pressure} shows the transport measurements for $x = 0.37$ under pressure. The sample remains in the AFM state for the pressure range we applied while the $\rho_{xx}(T)$ anomaly at $T_N$ moves to higher temperatures under pressure. As summarized in Fig. \ref{Pressure} (b), $T_N$ linearly increases with pressure at a rate of 1.4 K/GPa, which is smaller than 1.9 K/GPa for $x=0$ \cite{chen2019suppression}. Figure \ref{Pressure} (c) presents the pressure dependence of MR at 2 K. The $\sim 1\%$ drop in MR again suggests the ground state remains AFM. $H_s$ marked with the arrows increases gradually with increasing pressure. The pressure dependence of $H_s$ is summarized in Fig. \ref{Pressure} (d), which suggests a linear increase of $H_s$ at the rate of 1.8 T/GPa. Both the increase of $T_N$ and $H_s$ under pressure indicate that the external pressure enhances the AFM interlayer coupling, which is expected due to the decreasing lattice parameter $c$ under external pressure.

\subsection{Band topology engineering} 
To understand the evolution of band structures as a function of Pb doping, we performed first-principle calculations on the bulk band structures of (Mn$_{1-x}$Pb$_x$)Bi$_2$Te$_4$ using the generalized gradient approximation (GGA) plus correlation parameter U (GGA+U) with spin-orbit coupling (SOC). The results are summarized in Fig. \ref{DFT}. Our calculations reveal an insulating ground state for both MnBi$_2$Te$_4$ and PbBi$_2$Te$_4$, the two end materials. The orbital projection shows that the Bi-$p$ orbitals and the Te-$p$ orbitals dominate around the Fermi level (E$_F$), while the Mn-$d$ orbitals and the Pb-$p$ orbitals are far away from the E$_F$ (Fig. \ref{DFT} (a) and (b)). As shown in Fig. \ref{DFT} (a), for MnBi$_2$Te$_4$, there are clear band inversion features between the Bi-$p$ and Te-$p$ states at the $\Gamma$ point, supporting a magnetic topological insulator state which is consistent with the literature. For PbBi$_2$Te$_4$, contrary to MnBi$_2$Te$_4$ whose band inversion appears at the $\Gamma$ point, the band inversion of PbBi$_2$Te$_4$ occurs at the $Z$ point, resulting in a strong topological insulator phase due to the preservation of spatial inversion and time-reversal symmetry (Fig. \ref{DFT} (b)). Our calculation is consistent with the previous research on PbBi$_2$Te$_4$ \cite{menshchikova2011ternary, kuroda2012experimental}.

We now investigate the evolution of the band gap via the fine-tuning of the strength of SOC (Fig. \ref{DFT} (c)). We found that for both MnBi$_2$Te$_4$ and PbBi$_2$Te$_4$, the band gaps at the $\Gamma$ and $Z$ points decrease rapidly when increasing the strength of SOC. In particular, for MnBi$_2$Te$_4$, the bulk gap at the $\Gamma$ point first decreases to zero and then reopens as the SOC is larger than 75$\%$. On the other hand, for PbBi$_2$Te$_4$, we find the bulk gap at $Z$ point is the one that closes first and then reopens at SOC $\sim$ 60$\%$ (Fig. \ref{DFT} (d)). Therefore, topological phase transitions can appear when the SOC increases for both compounds.

Following the line of reasoning, will the Pb doping on MnBi$_2$Te$_4$ induce topological phase transitions? To shed light on this, we calculate the band structures of (Mn$_{1-x}$Pb$_x$)Bi$_2$Te$_4$. Figure \ref{DFT} (e) shows the minimum gap value between the valence band and conduction band as a function $x$. Generally, topological phase transition between magnetic and nonmagnetic states do not induce additional band inversion, because the two end of states possess different symmetry. However, this concept has its limitations, it is only valid to the trivial to nontrivial phase transition that occurs at the same time-reversal symmetry momenta in the two end systems. As we have shown that the band inversion in MnBi$_2$Te$_4$ and PbBi$_2$Te$_4$ appear at $\Gamma$ or $Z$ point, respectively. Thus complicated band inversion diagram is expected. Indeed, our results display two gapless points when doping concentration achieves $x = 0.44$ and $x = 0.66$. Since the band inversion may exist at $\Gamma$ and $Z$ simultaneously between these two ratios, we expect that there might be a new topological phase in this doping regime. Detailed DFT and angle-resolved photoemission spectroscopy (ARPES) study of the effect of doping in this material are left as an open question for future studies.

\section{Discussion}

\begin{figure}
    \centering
    \includegraphics[width=3.4in]{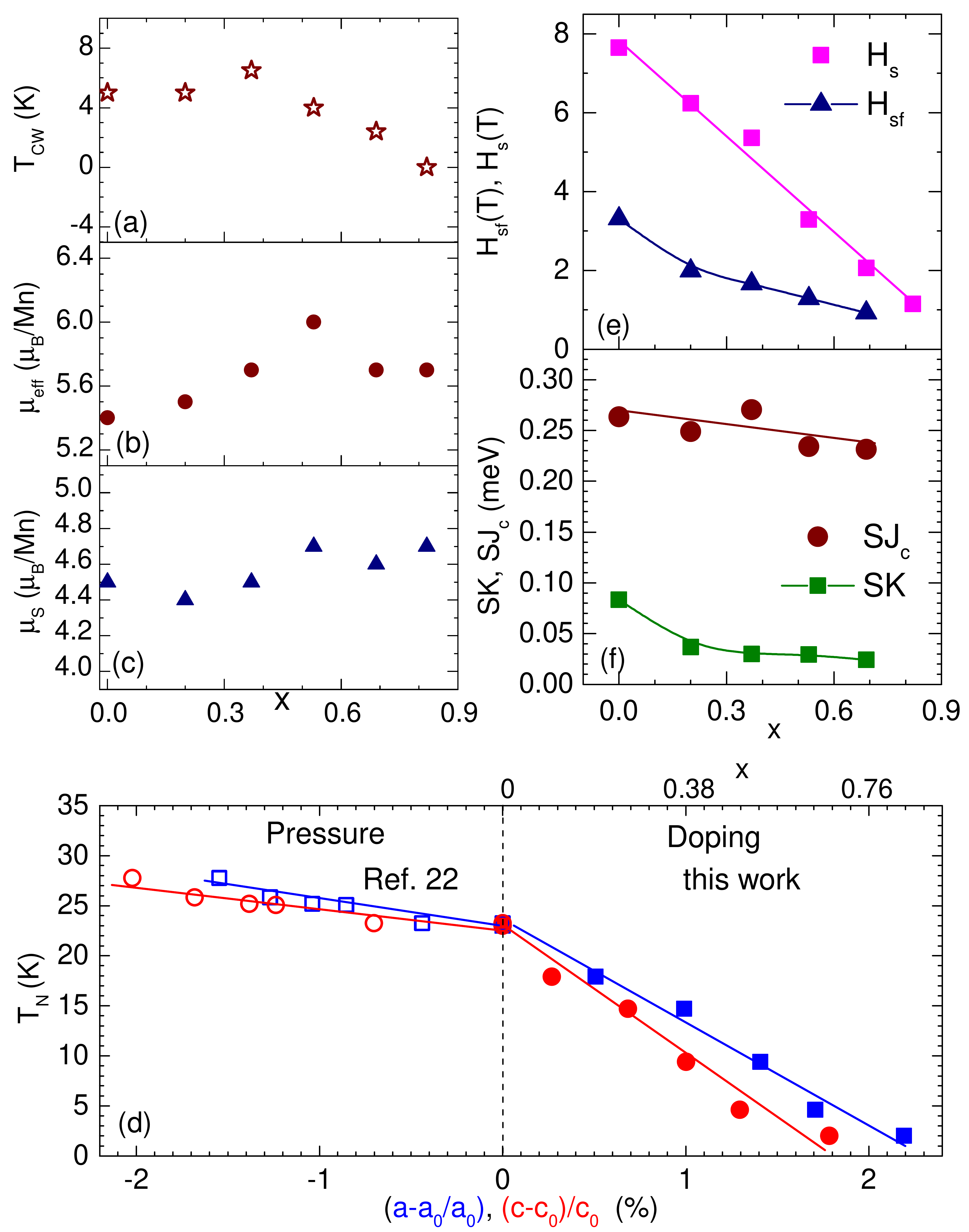}
    \caption{(a) The doping dependence of the Curie-Weiss temperature,
    (b) The doping dependence of the Curie-Weiss effective moment per Mn,
    (c) The doping dependence of the magnetic moment per Mn for $x=0$ (2 K, 7.7 T) and $x>0$ (2 K, 7 T),
    (d) N$\acute{e}$el temperature vs. $x$,  N$\acute{e}$el temperature vs. $(a-a_0)/a_0$ and N$\acute{e}$el temperature vs. $(c-c_0)/c_0$ for the pressure work (Ref. 22) and this doping work,
    (e) Spin-flop field and saturation field determined from magnetic and transport measurements with $H\parallel c$,
    (f) The doping dependence of the $effective$ (see the main text) interlayer plane-plane coupling per Mn site $SJ_c$ and the 
    $effective$ magnetic anisotropy per Mn $SK$. All lines are guides to the eye. }
    \label{summary} 
\end{figure}

Figures 7(a)-(f) summarize the doping-dependent magnetic properties. The doping-dependence and the magnitudes of $T_{CW}$ (Fig. 7(a)) are not trivial. The system is very 2D and one expects the $T_{CW}$ to be set by strong intraplanar ferromagnetic interactions
and scale with the average number of Mn neighbors, $i.e.$, as $1-x$, which is not the case here, especially in the Mn-rich side. We argue this is because at 40--250 K we may not be in the true Curie-Weiss regime due to the strong FM in-pane fluctuations, as indicated by the $\mu_{\rm{eff}}$ (Fig. 7(b)) being slightly smaller than the expected 5.9 $\mu_B$/Mn. Indeed, neutron scattering experiments indicate strong FM in-plane correlations even
at room temperature for MnBi$_2$Te$_4$\cite{Rob_unpublished}. Furthermore, in a 2D system where strong fluctuation always exists, one would expect 
the $T_N$ to be strongly suppressed compared to the mean-field limit value, $T_\mathrm{MFT}$; indeed, even in the least-fluctuating square Ising
model, $T_N$ is nearly smaller than half of $T_\mathrm{MFT}$. On the contrary, $T_{CW}$ we obtained using the Curie-Weiss fit of our data for 40-250 K listed in Table I is much smaller than $T_N$. This may be partially because we are not in the true Curie-Weiss regime as aforementioned. But interestingly, similarly odd 
behavior was observed in some other quasi-2D ferromagnet or A-type antiferromagnets. For instance, $T_N=14$ K and high-temperature $T_{CW}=11(1)$ K for CrCl$_3$ \cite{mcguire2017magnetic}; $T_N=61$ K and $T_{CW}=71(1)$ K for CrI$_3$ \cite{mcguire2015coupling}. These large $T_N$/$T_{CW}$ ratio, to the best of our knowledge, was never explained, since a quantitative theory of spin-susceptibility in Mermin-Wagner systems has never been worked out.

Upon doping, $\mu_s$ per Mn slightly increases (Fig. 7(c)), suggesting possible reduction in the number of the Mn$_{\rm{Bi}}$ antisites, consistent with our neutron scattering refinement. Figure 7(d) shows a comparison of the pressure work \cite{chen2019suppression} and our doping work. Apparently, $dT_N/da$ or $dT_N/dc$ is much larger in Sb-doped MnBi$_2$Te$_4$ than that in the pressurized MnBi$_2$Te$_4$. This is reasonable since the former comes from both the magnetic dilution and lattice expansion while the latter is only caused by lattice expansion.
Furthermore, comparing with (Mn$_{1-x}$Sn$_x$)Bi$_2$Te$_4$ \cite{zhu2021magnetic} where $T_N \sim 18$ K and $H_s \sim 6$ T at $x=0.5$, Pb doping shows a much stronger suppression of magnetism with $T_N\sim 9$ K and $H_s\sim 3$ T at $x=0.5$ (Fig. 7(d) and (e)). Due to the larger atomic radius difference between Pb and Mn, Pb doping can cause a faster lattice parameter increase than Sn, it is thus reasonable to expect a faster suppression of the AFM coupling, $T_N$ and $H_s$. 

Unlike the monotonic decrease of $T_N$ and $H_s$ in Sb-doped MnBi$_2$Te$_4$ where the Mn1 sublattice is diluted and the Mn2 sublattice gets enhanced upon doping, here $T_N$ and $H_s$ decrease rather linearly as shown in Fig. \ref{summary}(d) and (e), leading to $T_N= 24-27.4x$ and $H_s=7.70-7.98 x$ up to $x=0.82$. The clear difference between these two doping series indicates that indeed the Pb-doping series is ideal to investigate the unadulterated magnetic dilution effect in MnBi$_2$Te$_4$. 

So now let us understand these behaviors accounting for the dilution effect when a non-magnetic Pb replaces a magnetic Mn so that the fraction of magnetic site is $\delta=1-x$. In a quasi-2D AFM system, long-range order is impossible without either interlayer coupling, $J_c$, or uniaxial
magnetic anisotropy, $K$. We define the former as the effective coupling strength between two neighboring
planes per Mn site and the latter as magnetic anisotropy parameter per Mn site. That is to say, the effective 
interplanar coupling $J_c$ includes all possible Mn-Mn exchange paths between the planes, and the effective anisotropy $K$ includes both single-ion and exchange anisotropies. 

We can write the full Hamiltonian as:
\begin{equation}
E = E_0+\delta^2J_c{\bf S}_i\cdot{\bf S}_{i+1}-\delta K(S_i^z)^2-\delta g\mu_B{\bf S}_i\cdot{\bf H},
\label{b1}
\end{equation}
where $g$ is the Lande factor, $i$ labels Mn planes, $J_c>0$ for AFM and $K>0$ to ensure $z$ is the easy axis. Since a magnetic bond needs to have Mn on both ends, $\delta^2$ arises for the two-site exchange term. Meanwhile, $\delta$ arises for the single-ion magnetic anisotropy and Zeeman terms. Thus the energy in the spin-flop phase is \cite{supplement},
\begin{align}
E(H,\phi)=&(E_1\mp \delta KS^2/2)+\delta^2J_c S^2 \cos(\pi-2 \phi) 
\nonumber\\
&\pm \delta K S^2 \cos^2\phi 
- \delta g \mu_B S H \sin\phi,\label{b}
\end{align}
where $E_1=E_0-\delta KS^2/2$, the upper (lower) sign corresponds to the angle between $\bf H$ and $\bf S$ as $\pi/2-\phi$ with $H$ along the easy axis $c$ ($H$ along the hard $ab$ plane).
By minimizing Eq.\ref{b} at $\phi=\pi/2$, we can get the saturation fields, 
\begin{align}
H_s^{\parallel c} &= 2(2\delta J_c-K)S/g\mu_B\label{ss}\\
H_s^{\parallel ab} &= 2(2\delta J_c+K)S/g\mu_B,
\end{align}
Similarly, one can estimate the spin-flop threshold:
\begin{equation}
    H_{sf}=\sqrt{K(2\delta J_c-K)}(2S/g\mu_B), \label{s}
\end{equation}
from Eq.\ref{ss} and \ref{s}:
\begin{align}
SK&=(g\mu_B/2)(H_{sf}^2/H_s^{\parallel c})\\
SJ_c &=(g\mu_B/4\delta)\left(H_s^{\parallel c}+H_{sf}^2\right/H_s^{\parallel c}).
\end{align}
Note that this scaling is only true if magnetic anisotropy is of a single-ion origin. If there is a contribution from the exchange anisotropy, that contribution will be scaled as $\delta^2$, and our $K$ in Eq.4-6 will be replaced as $ K_1+\delta K_2 $ where $K_1$ is the single-ion anisotropy parameter and $K_2$ is the exchange anisotropy parameter.

Using Eq.7 and 8, we estimate the effective $SK$ and $SJ_c$, as shown in Fig. \ref{summary} (f). $SJ_c$ slightly decreases from 0.26 meV at $x=0$ to 0.23 meV at $x=0.69$,
being consistent with the small change in lattice parameter $c$. Meanwhile $SK$ shows a monotonic decrease with a sharp drop from 0.08 meV at $x=0$ to 0.04 meV at $x=0.20$ and then a slow decrease to 0.02 meV at $x=0.69$. We thus readily see that $H_s$ is linear in $\delta$, as seen in Fig. \ref{summary} (e), because it is defined mostly by $\delta J_c$. But the behavior of $ H_{sf}$ is harder to understand:
naively, it can behave either sublinearly, or, in the extreme case of the dominating exchange anisotropy, linearly with $\delta$. Figure \ref{summary} (e) shows that for $x\agt 0.2$ the behavior is indeed linear, suggesting that the anisotropy there is dominated by the exchange anisotropy. But there is an additional contribution at $x=0$, of about 0.05 meV, which mostly disappears at $x=0.2$. The only plausible explanation is that
this contribution comes from the single-ion anisotropy which is strongly affected by the local environment, and only appears if all or nearly all of the nearest neighbors
of a given Mn ion are also Mn. It is easy to see that the probability of having a Mn at a
given site, $and$ having all its neighbors Mn, is $\delta^7$, and is only 100\% at $x=0$, 20\% at $x=0.2$, and 8\% at $x=0.3$.

We can try to understand the linear doping dependence of $T_N$ by studying the magnetic dilution effect in the mean-field limit ($i.e.$, in the Weiss
molecular field theory). We consider an individual Mn ion with the spin $S$ and 6 nearest sites. Under doping, the mean-field-theory temperature $T_{\mathrm {MFT}}$ (we use this
notation to distinguish it from the $T_{CW}$ extracted experimentally 
from $1/\chi(T)$ for 40--250 K, which, as discussed above, does not represent the true MFT limit) is given by $
T_{\mathrm {MFT}} \propto \delta \mu_{eff}^{2},
$
which linearly decreases with $x$. Given that $T_N$ is, generally speaking, nothing but
fluctuations-renormalized mean-field-theory temperature, $T_{N}\approx
T_\mathrm{MFT}/(a+b\log({\bar{J}}/{\bar{J}_{c}})),$ where $a$ and $b$ are not
supposed to change much with doping, $J$ is the intraplanar magnetic coupling and $J\gg J_c$, bars means spacial average. As discussed above, $\bar{J}\sim \delta^{2}J,$ and
$\bar{J}_{c}\sim \delta^{2}J_{c},$ so $\log(\bar{J}/\bar{J}_{c}) \sim \log({{J}}/{{J}_{c}})$. Since $T_N$ depends on $J$ and $J_c$ logarithmically weakly, we conclude that $T_{N}$ should roughly follow $T_\mathrm{MFT}$ and thus linearly decreases with $x$.

Lastly, the bottom row of Fig. \Ref{H-dependence} shows that a sign change of the anomalous Hall resistivity $\rho_{xy}^A(H)$ occurs between $x=0.37$ and $x=0.53$, may suggesting possible band structure changed in this regime. Thus we call ARPES experiments to investigate the band structures of this doping series to address this question.

\section{Conclusion} 

In summary, we have grown high-quality single crystals of (Mn$_{1-x}$Pb$_x$)Bi$_2$Te$_4$ with $x$ ranging from 0 to 0.82. We find that this doping series provides a great platform to investigate the magnetic dilution effect in van der Waals magnets. The N$\acute{e}$el temperature and saturation field decrease linearly with doping, which can be well understood in a simple model considering the dilution effects. Moreover, our DFT calculations reveal two gapless points appearing at $x = 0.44$ and $x = 0.66$. Together with the sign change of the anomalous Hall resistivity between $x=0.37$ and $x=0.53$, this may suggest possible
topological phase transitions in this doping series.

\section*{Acknowledgments}
We thank  Robert J. McQueeney and Steven Winter for useful discussions. Work at UCLA was supported by the U.S. Department of Energy (DOE), Office of Science, Office of Basic Energy Sciences under Award Number DE-SC0021117. T.-R.C. was supported by the Young Scholar Fellowship Program from the Ministry of Science and Technology (MOST) in Taiwan, under a MOST grant for the Columbus Program MOST110-2636-M-006-016, NCKU, Taiwan, and National Center for Theoretical Sciences, Taiwan. Work at NCKU was supported by the MOST, Taiwan, under grant MOST107-2627-E-006-001 and Higher Education Sprout Project, Ministry of Education to the Headquarters of University Advancement at NCKU. Work at ORNL was supported by US DOE BES Early Career Award KC0402010 under Contract DE-AC05-00OR22725 and used resources at the Spallation Neutron Source and the High Flux Isotope Reactor, DOE Office of Science User Facilities operated by the Oak Ridge National Laboratory. I. M. acknowledges support from DOE under the grant DE-SC0021089. C. H. thanks was supported by the Julian Schwinger Fellowship at UCLA.

\medskip

\bibliographystyle{apsrev4-1}
\bibliography{PbMBTbib}

\end{document}